\documentclass[aps,pra,twocolumn,groupedaddress,showpacs]{revtex4-1}
\usepackage{graphicx}
\usepackage{amsmath}
\usepackage{amsfonts}
\usepackage{amssymb}
\usepackage[normalem]{ulem}
\usepackage{xcolor}
\usepackage{float}

\def\abs#1{{\left|#1\right|}}

\def\tFWHM{\tau_\text{FWHM}}
\def\tctrl{\tau^\text{ctrl}_\text{FWHM}}
\def\Deltau{\Delta\tau^\text{ctrl}}
\def\GG{\mathcal{G}_G}

\newcommand{\overbar}[1]{\mkern 7mu\overline{\mkern-7mu#1\mkern-3mu}\mkern 3mu}

\definecolor{darkgreen}{RGB}{34,139,34}

\begin{document}

\title{Variance-Based Sensitivity Analysis of $\Lambda$-type Quantum Memory}
\author{Kai Shinbrough$^{1,2}$}
\email{kais@illinois.edu}
\author{Virginia O. Lorenz$^{1,2}$}

\address{$^1$Department of Physics, University of Illinois at Urbana-Champaign, 1110 West Green Street, Urbana, IL 61801, USA\\
$^2$Illinois Quantum Information Science and Technology (IQUIST) Center, University of Illinois at Urbana-Champaign, 1101 West Springfield Avenue, Urbana, IL 61801, USA}

\date{\today}

\begin{abstract}
The storage and retrieval of photonic quantum states, quantum memory, is a key resource for a wide range of quantum applications. 
   Here we investigate the sensitivity of $\Lambda$-type quantum memory to experimental fluctuations and drift. We use a variance-based approach, focusing on the effects of fluctuations and drift on memory efficiency.  
   We consider shot-to-shot fluctuations of the memory parameters, and separately we consider longer timescale drift of the control field parameters. We find the parameters that a quantum memory is most sensitive to depend on the quantum memory protocol being employed, where the observed sensitivity agrees with physical interpretation of the protocols. We also present a general framework that is applicable to other figures of merit beyond memory efficiency. These results have practical ramifications for quantum memory experiments.
\end{abstract}

\maketitle

\section{Introduction}

In the emerging field of quantum technology, photons play a critical role as carriers of quantum information \cite{BB84,Sangouard11,DLCZ} and as the fundamental qubits for quantum computation and information processing \cite{KLM,Raussendorf01}. Photons are, however, difficult to synchronize \cite{Nunn13,Kaneda19,Makino16} and are subject to losses in transmission \cite{Sangouard08,Sangouard11,Briegel98,Teleportation}. The ability to store and retrieve photonic quantum states on demand---quantum memory---provides a path forward to overcome these challenges, and is therefore a critical enabling technology for future quantum applications \cite{Lvovsky09,Simon10,Ma17}. A considerable body of work has been dedicated to quantum memories based on atomic ensembles, where the three-level, resonant, $\Lambda$-type atomic system is the most common \cite{Lvovsky09,Simon10,Ma17,GorshkovPRL,GorshkovI,GorshkovII,GorshkovIII,GorshkovIV,Nunn07,Fleischhauer02,ATS,Vivoli13,Moiseev01}.

In the ideal case, an optical quantum memory is capable of storing single-photon quantum states and retrieving them on demand with high efficiency, high fidelity, long storage time, and broad bandwidth \cite{Lvovsky09,Simon10,Ma17}. Another critical indicator of quantum memory performance, however---which has largely been neglected until only recently \cite{Teja21,Otten21}---is a memory's sensitivity to experimental fluctuations and drift. 
Fluctuations and drift in experimental parameters are invariably present in physical quantum memory implementations, and a memory which is more robust (less sensitive) to experimental noise is more useful for real-world quantum applications. 
Here we quantitatively address this aspect 
of $\Lambda$-type quantum memory
. (For an analysis of other types of memories, we refer the reader to Refs.~\cite{Teja21,Otten21}.) We provide a variance-based sensitivity analysis \cite{Cacuci03book,Castillo08,Saltelli10,Sobol01,Sobol93,Pearson1905,Saltelli19,saltelli08Book}, which sheds light on not only the sensitivity of an individual quantum memory implementation with device-specific fluctuations and drift, but also on the intrinsic sensitivity of different physical $\Lambda$-type quantum memory protocols.

\begin{figure}[t]
	\centering
	\includegraphics[width=\columnwidth]{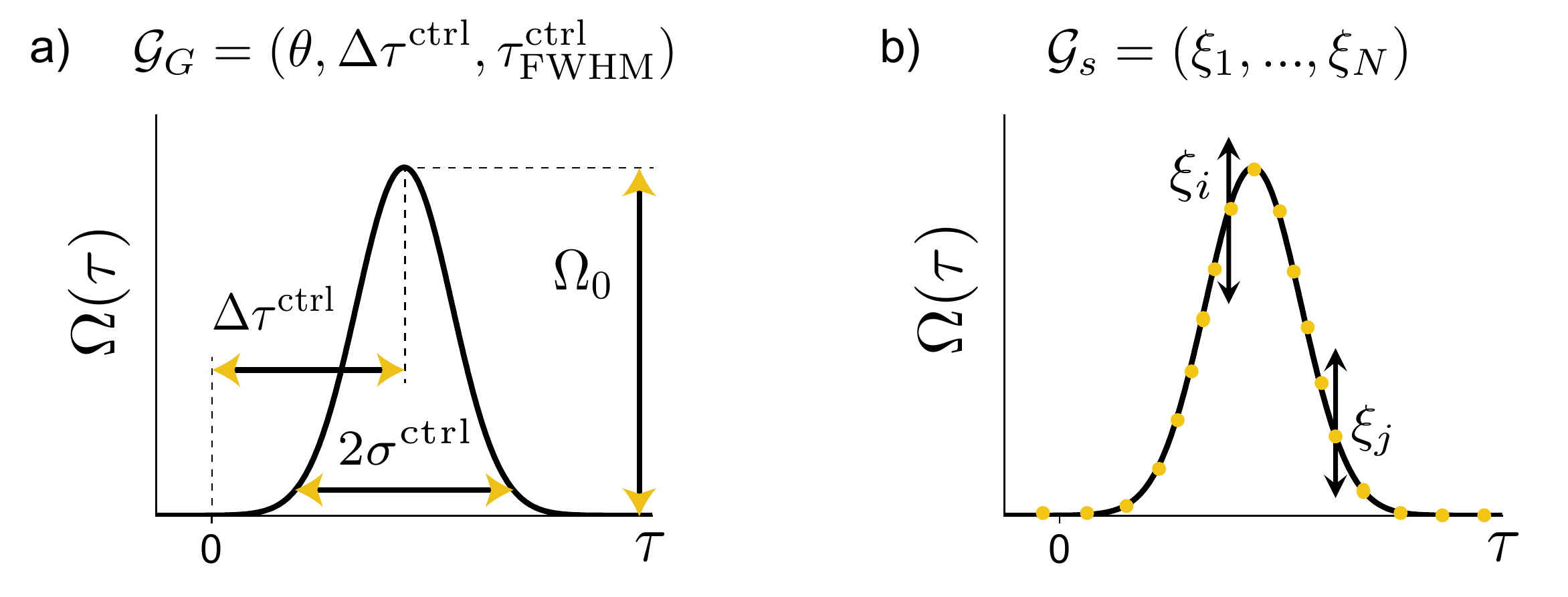}
	\caption{Control fields of (a) Gaussian shape, defined by the three parameters $\GG = (\theta,\Deltau,\tctrl)$, where $\Omega_0 = \theta/(2\sqrt{\pi}\sigma^\text{ctrl})$ and $\sigma^\text{ctrl}=2\sqrt{2\ln{2}}/\tctrl$, and (b) arbitrary shape, defined by the $N$ interpolation points $\mathcal{G}_s = (\xi_1,...,\xi_N)$.}
	\label{Fig_Gauss_v_Shape}
\end{figure}

We consider quantum memory implementations with 
memory parameters $\mathcal{M}=(d,\tFWHM\gamma)$, where the optical depth $d$ of the atomic ensemble and the intermediate state coherence decay rate $\gamma$, scaled by the signal photon duration $\tFWHM$, are considered to be intrinsic and fixed properties of the memory. We then group the remaining extrinsic, more readily tunable parameters as $\mathcal{G}$, which parameterize the optical control field used in the memory interaction, and which we assume have been optimized in order to maximize memory efficiency. We further partition our analysis according to whether the parameters of the control field define a Gaussian temporal envelope $\GG = (\theta,\Deltau,\tctrl)$ or an arbitrary temporal shape $\mathcal{G}_s = (\xi_1,...,\xi_N)$, as investigated in Ref.~\cite{GaussianPaper} (see Figure~\ref{Fig_Gauss_v_Shape}), where $\theta$, $\Deltau$, and $\tctrl$ correspond to the Gaussian control field pulse area, delay relative to the signal field, and duration, respectively, and the points $\xi_i$, $i=1,...,N$ correspond to interpolation points along the temporal envelope of the control field.  Details on the numerical calculation of memory efficiency given $\mathcal{M}$ and $\mathcal{G}$ can be found in Ref.~\cite{GaussianPaper}. 

In this work we assume a typical scenario for experimental atomic-ensemble quantum memory, wherein the memory parameters are fixed with minimal long-timescale drift at a given setpoint 
but may undergo non-negligible shot-to-shot fluctuations. This situation occurs frequently in transient processes for generating dense atomic ensembles, such as in light-induced atomic desorption (LIAD) \cite{LIAD1,LIAD2} or laser ablation, but applies to equilibrium systems as well. We assume the optical parameters of the control field possess smaller shot-to-shot fluctuations (e.g., laser fields with locked frequency, power, timing, etc.), but may either drift over time or may not be set precisely for optimal memory performance. We investigate the sensitivity of the memory performance to the setting of these control parameters, including analysis of correlations that exist between parameters, which may allow, for example, for compensating a drop in efficiency due to non-optimal setting of one parameter by modification of the remaining parameters. This latter analysis may be important in situations where one parameter is constrained experimentally, for example in the case of limited laser power, which can often limit memory efficiency \cite{Reim10,England13,Bustard13,England15,Michelberger15,Fisher17,BinBaMemory,Ding15}. This type of memory sensitivity can be interpreted as an indicator of the region of control field phase space where acceptable memory performance can be achieved; low sensitivity implies a large acceptable region of control field phase space, where the control field does not require careful fine-tuning, and where restrictions on one parameter may be compensated for with changes to the remaining parameters. Equivalently, this type of memory sensitivity 
can be interpreted in terms of the memory's robustness to experimental drift, where low sensitivity implies that, given optimal initial control field settings, the memory will be robust to long-timescale drift in the phase space surrounding the optimal setpoint.

In the following sections, we restrict our discussion to resonant $\Lambda$-type memory protocols, but the tools developed in this work are readily applicable to off-resonant protocols, as well as other level systems and a wide range of related techniques \cite{ORCA,AFC1,SensorProtection}. In Section~\ref{DefSec}, we provide definitions for several quantitative aspects of memory sensitivity. In Section~\ref{FlucSec} we use these criteria to analyze the sensitivity of resonant $\Lambda$-type quantum memory to fluctuations in memory parameters, and in Section~\ref{DriftSec} we address sensitivity to improper setting of control field parameters or experimental drift.

\section{Variance-Based Sensitivity Analysis}\label{DefSec}

The sensitivity of classical 
systems is a much-discussed subject with well-established theoretical and numerical tools \cite{Cacuci03book,Castillo08,Saltelli10,Sobol01,Sobol93,Sobol05,Pearson1905,Saltelli19,saltelli08Book}. In general, the task is to determine the sensitivity of a system with performance criterion $h(\mathcal{X},\mathcal{A})$ to changes in $N$ input parameters $\mathcal{X}=(x_1,...,x_N)$ when internal system parameters $\mathcal{A}$ are kept fixed. 
This performance criterion may correspond to any desired single-valued metric of the system; in the case of quantum memory, this may correspond to memory efficiency, fidelity, storage time, etc. For the sake of brevity, in Sec.~\ref{FlucSec} and Sec.~\ref{DriftSec} we focus on memory efficiency as a key performance criterion, but importantly other criteria may be used and may be the subject of future work. In this section, we provide an outline of the theoretical tools used for a generic criterion $h$. 

The most common method for determining the sensitivity of $h(\mathcal{X},\mathcal{A})$ to fluctuations in the input parameters proceeds as follows 
\cite{Cacuci03book,saltelli08Book}. We define center values for the input parameters $\overline{\mathcal{X}}$, 
then draw many $N$-dimensional fluctuations $\zeta$ stochastically from a known probability distribution $P(\zeta)$, and average over these fluctuations in order to calculate the mean performance criterion 

\begin{equation}\label{etabarfluc}
    \overline{h}(\overline{\mathcal{X}}) = \int d\zeta\, h(\overline{\mathcal{X}}+\zeta,\mathcal{A}) P(\zeta)
\end{equation}

\noindent and the variance in the system performance 

\begin{equation}\label{Vfluc}
    V^\text{fluc}_h\left(\overline{\mathcal{X}}\right) = V_{\zeta}[h(\overline{\mathcal{X}}+\zeta,\mathcal{A})|\mathcal{A}],
\end{equation}

\noindent where $V_x[y(x,z)|z] = \int dx\, y^2(x,z)P(x) - [\int dx\, y(x,z)P(x) ]^2$ is the unconditional variance of $y$ obtained when $x$ is allowed to vary and $z$ is held constant. In the absence of a tailored noise model, the probability distribution for fluctuations is commonly approximated as an $N$-dimensional normal distribution $P(\zeta)\sim e^{-\abs{\zeta}^2/(2\epsilon^2)}$ with standard deviation $\epsilon$. The resulting standard deviation in performance criterion $h$ can then be calculated, $\sigma_h^\text{fluc}(\overline{\mathcal{X}}) = \sqrt{V^\text{fluc}_h(\overline{\mathcal{X}})}$.  

The simple variance-based method above provides useful information on the response of the system to short-timescale, shot-to-shot fluctuations in input parameters around given central values $\overline{\mathcal{X}}$, which typically correspond to the setpoints of the input parameters. $\mathcal{X}$ can also correspond to control parameters, where the setpoint $\overline{\mathcal{X}}$ is assumed to be at or near the optimum values for system performance. The method above does not provide detailed information on the local environment around the performance optimum, which may be important for long-timescale drift 
or for determining which parameter is most sensitive to experimental error. The simplest method for determining a system's sensitivity to these long-timescale changes in input parameters $\mathcal{X}=(x_1,...,x_N)$ is to vary each parameter one-at-a-time (OAT), and to measure the resulting variance in the system's performance. This OAT analysis corresponds to calculating the variances

\begin{equation}\label{OAT_var}
    V_{i}^\text{OAT} = V_{x_i}[h(\mathcal{X})\vert x_{j\neq i}]
\end{equation}

\noindent for each parameter $x_i$, where $x_i$ varies over a finite range, $x_i\in[x_i^\text{min},x_i^\text{max}]$. In Eq.~\eqref{OAT_var} and in the following discussion, we have suppressed the internal parameters that are always held constant from the notation
. Again, the standard deviation $\sigma_i^\text{OAT} = \sqrt{V_i^\text{OAT}}$ may be used to quantify the change in system performance due to parameter $x_i$. The parameter $x_i$ with the largest $\sigma_i^\text{OAT}$ has the largest effect on the performance criterion $h$ and therefore the largest sensitivity. In practice, this means stabilizing and optimizing that parameter 
is the most important for system performance and should receive the largest dedication of resources. 



\begin{figure}[t]
	\centering
	\includegraphics[width=0.55\columnwidth]{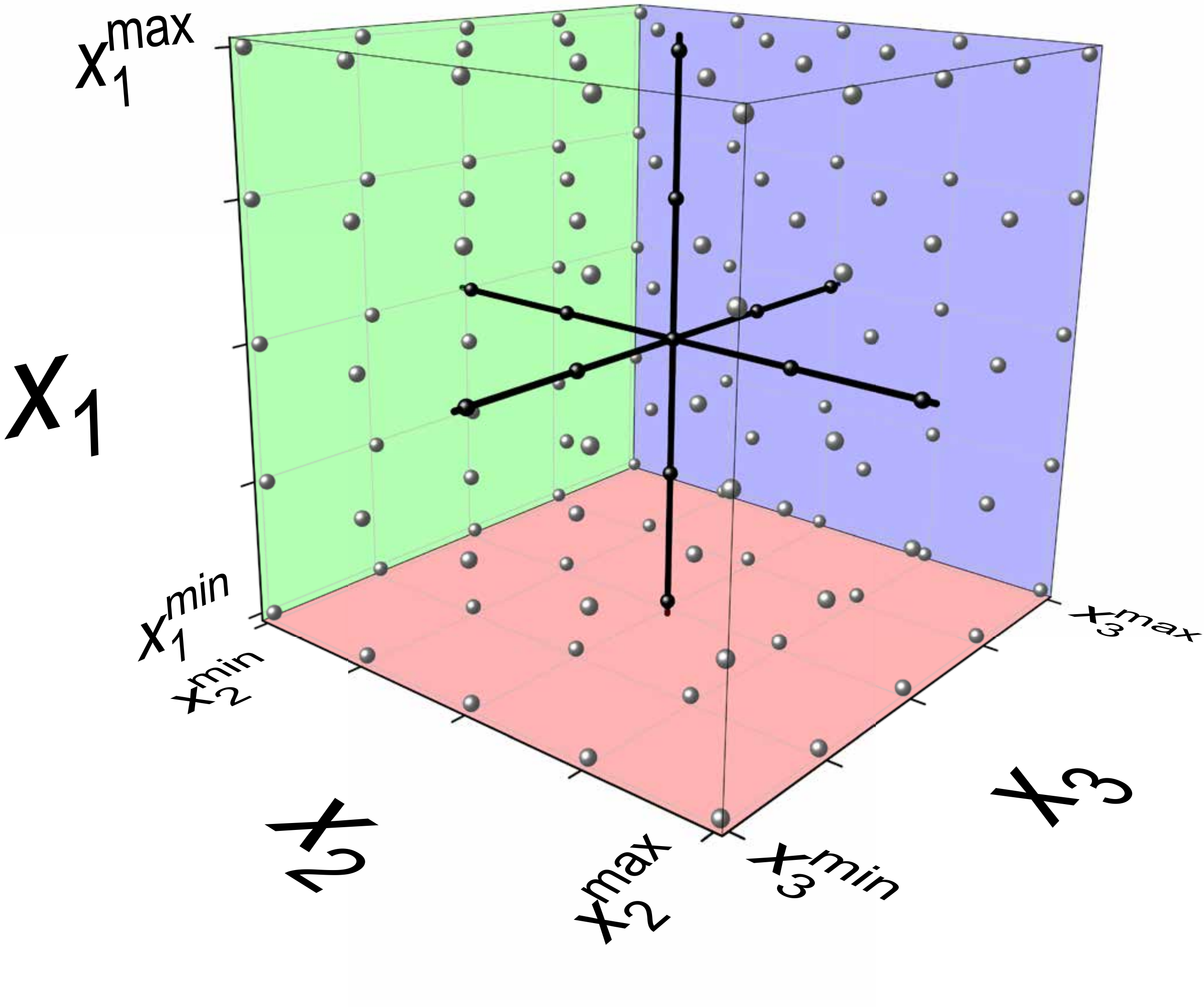}
	\caption{Example of 3-dimensional phase space spanned by parameters $\mathcal{X}=(x_1,x_2,x_3)$; the sample points connected by the black lines correspond to one-at-a-time (OAT) analysis of the system sensitivity, which explores only a small fraction of the total phase space.
	}
	\label{Fig_PhaseSpace}
\end{figure}

The OAT analysis above provides cross-sectional information on the local environment around a performance optimum, and can be used to rapidly determine if one input parameter is responsible for the majority of observed variance in system performance. As correlations between input parameters arise and the dimensionality of $\mathcal{X}$ increases, however, OAT analysis rapidly becomes insufficient as it ignores control parameter correlations and explores only a small fraction of the input phase space (see Figure~\ref{Fig_PhaseSpace}) \cite{saltelli08Book,Saltelli19}. When correlations between parameters exist or the dimensionality of $\mathcal{X}$ is large, \textit{global} variance-based sensitivity analysis is required, wherein the most prevalent sensitivity measure is the first-order Sobol' variance \cite{Sobol01,Sobol05,Sobol93} 

\begin{equation}\label{Vi}
    V_i = V_{x_i}\{E[h(\mathcal{X})\vert x_i]\},
\end{equation}

\noindent where the inner expectation value, $E[\cdot]$, corresponds to the mean of $h(\mathcal{X})$ when $\mathcal{X}$ is varied over all possible values in a finite range at fixed $x_i$. The outer variance then measures the variance of this mean with respect to changes in $x_i$. For a 3-dimensional parameter space, shown in Fig.~\ref{Fig_PhaseSpace}, the variance $V_1$ in Eq.~\eqref{Vi} corresponds to calculating the expectation value $E[h(\mathcal{X})\vert x_1]$ over each red plane where $x_1$ is fixed and $x_2$ and $x_3$ are allowed to vary, and taking the variance of these expectation values. Similarly, $V_2$ ($V_3$) corresponds to calculating the variance of expectation values taken over the green (blue) planes. This method fully explores the parameter phase space in Fig.~\ref{Fig_PhaseSpace}, whereas OAT analysis only explores the small region of parameter space connected by the black lines. The first-order Sobol' sensitivity index can be calculated as

\begin{equation}\label{Si}
	S_i = V_i/V_\text{tot},
\end{equation}

\noindent where $V_\text{tot}$ is the total variance $V_\mathcal{X}[h(\mathcal{X})]$ observed over the range of interest. Importantly, this technique also allows for the calculation of higher-order variances and sensitivities:

\begin{align}
	V_{ij} &= V_{x_i,x_j}\{E[\eta(\mathcal{X})\vert x_i,x_j]\} - V_i - V_j,\\ 
	S_{ij} &= V_{ij}/V_\text{tot},
\end{align}

\noindent which are related with Eqs.~\eqref{Vi} and \eqref{Si} by the conditions

\begin{align}
	V_\text{tot} &= \sum_i V_i + \sum_i\sum_{j>i} V_{ij} + ... + V_{1...N}, \\
	S_\text{tot} &= \sum_i S_i + \sum_i\sum_{j>i} S_{ij} + ... + S_{1...N} = 1,
\end{align}

\noindent and which probe correlations between parameters. In 3 dimensions, these higher order variances correspond to the variance of expectation values evaluated along lines 
in the phase space of Fig.~\ref{Fig_PhaseSpace}, instead of planes.

Sobol' variances and sensitivity indices provide a complete picture of the system performance landscape around a central point of input parameters, and allow for identification of which input parameters are most sensitive globally. This analysis also probes whether correlations exist between parameters, which can be leveraged to allow for acceptable system performance at non-optimal parameter values.

\section{Fluctuations in Resonant $\Lambda$-type Quantum Memory}\label{FlucSec}

We now apply the general discussion in Sec.~\ref{DefSec} to the case of resonant, $\Lambda$-type quantum memory, beginning with the effect of fluctuating memory parameters on memory efficiency.  In the resonant case, there exist three well-known quantum memory protocols: the electromagnetically induced transparency (EIT) \cite{Fleischhauer02,Phillips01,Liu01,Gorshkov07_2}, Autler-Townes Splitting (ATS) \cite{ATS,ATS2,ATS3,ATS4}, and absorb-then-transfer (ATT) \cite{Moiseev01,Vivoli13,Gorshkov07_2,Carvalho20,GaussianPaper} protocols.

\begin{figure*}[t]
	\centering
	\includegraphics[width=\linewidth]{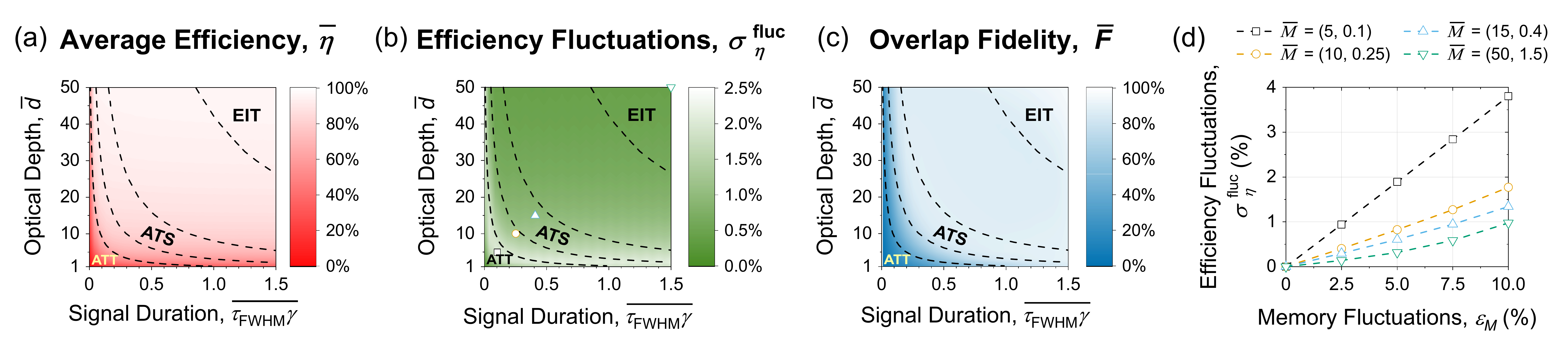}
	\caption{Sensitivity of $\Lambda$-type quantum memory to fluctuations $\epsilon_M$ in memory parameters around the center points $\overbar{\mathcal{M}} = (\overline{d},\overline{\tau_{\text{FWHM}}\gamma})$. (a) Average efficiency $\overline{\eta}$ and (b) fluctuations in memory efficiency $\sigma_\eta^\text{fluc}$ in the presence of fluctuations $\epsilon_M=5\%$. (c) Average overlap fidelity $\overline{F}$ of optimal Gaussian control field parameters. (d) Fluctuations in memory efficiency as a function of increasing magnitude of fluctuations in memory parameters for selected points shown in (b). Regions for ATT, ATS, and EIT memory protocols are enclosed with dotted lines.}
	\label{Fig_Fluctuations}
\end{figure*}

We consider the case where fluctuations in 
memory efficiency are dominated by fluctuations in the input parameters $\mathcal{M}$, and we assume the control field parameters $\mathcal{G}$ are kept fixed at the optimum values for the average memory parameters $\overbar{\mathcal{M}}$ (i.e., the memory parameter setpoints). We assume a generic noise model where fluctuations $\zeta_M = (\zeta_{d},\zeta_{g})$ are drawn stochastically from the probability distribution 
%
    $P(\zeta) \sim e^{-(\zeta_d^2 g^2 + \zeta_g^2 d^2)/[2(\epsilon_M dg)^2]}$ using the Mersenne Twister algorithm with seed 0, where $g=\tau_\text{FWHM}\gamma$. 
This implies that, e.g., for $\epsilon_M=5\%$, both memory parameters vary by 5\% of their respective setpoints. We further correlate optical depth and linewidth so as to preserve atom number. We calculate both $\overline{\eta}(\overbar{\mathcal{M}})$ and $\sigma_\eta^\text{fluc}(\overbar{M})$ following the prescription of Eqs.~\eqref{etabarfluc} and \eqref{Vfluc}, assuming the optimal Gaussian control field values $\GG(\overbar{\mathcal{M}})$ found in Ref.~\cite{GaussianPaper}. For each center value $\overbar{\mathcal{M}}$ we average over 1000 random fluctuations $\zeta_M$. 

The results of this analysis are shown in Fig.~\ref{Fig_Fluctuations}(a) and (b) for $\epsilon_M=5\%$, where we have labeled the respective regions of memory parameter space for optimal ATT, ATS, and EIT quantum memory protocols using the same procedure outlined in Ref.~\cite{GaussianPaper}. We find that the largest fluctuations in memory efficiency occur in the region of memory parameter space below the ATS regime, in the so-called absorb-then-transfer (ATT) regime. 
The ATS protocol is much less sensitive to memory parameter fluctuations, as $\sigma_\eta^\text{fluc}$ is reduced by approximately a factor of 2, but in turn the EIT protocol is approximately a factor of 2 less sensitive to these fluctuations than the ATS protocol. 

To explain this behavior physically, we consider the changes in optimized control field parameters $\GG$ as a function of $\mathcal{M}$. As shown in Ref.~\cite{GaussianPaper}, the gradient of $\GG$ with respect to changes in $\mathcal{M}$ is largest in the non-adiabatic ($d\tau_\text{FWHM}\gamma\lesssim 1$) regime and becomes smaller as the memory adiabaticity increases. This implies that, if the memory protocol is non-adiabatic, the optimal parameters $\GG$ change significantly even for small changes in $\mathcal{M}$ and thus these small changes may cause comparatively large changes in memory efficiency compared to the adiabatic regime. This intuition can be evaluated quantitatively by considering the average overlap fidelity of $\mathcal{G}=\GG(\mathcal{M})$ and $\mathcal{G}'=\GG(\mathcal{M}')$ at different memory parameters $\mathcal{M}'=\mathcal{M}+m$:

\begin{equation}\label{Fbar}
    \overline{F}(\mathcal{M}) = \frac{1}{A}\int_0^R dm^2 \, F(\mathcal{G},\mathcal{G}'),
\end{equation}

\noindent where $m$ varies over a 2D region with radius $R$ and area $A$, and the overlap fidelity between any two points is

\begin{equation}
    F(\mathcal{G},\mathcal{G}') = \frac{\abs{\int_{-\infty}^{\infty}d\tau \, \Omega^*(\mathcal{G})\Omega(\mathcal{G}')}^2}{\int_{-\infty}^{\infty} d\tau \, \abs{\Omega(\mathcal{G})}^2 \int_{-\infty}^{\infty} d\tau \, \abs{\Omega(\mathcal{G}')}^2}.
\end{equation}

\noindent The control field $\Omega(\mathcal{G})$ is a Gaussian function with area $\theta$, duration $\tctrl$, and timing $\Deltau$. The average overlap fidelity in Eq.~\eqref{Fbar} is shown in Fig.~\ref{Fig_Fluctuations}(c), and confirms the intuition that the region of least overlap corresponds to the absorb-then-transfer protocol, where the memory parameters are most non-adiabatic. It is therefore this region that is most sensitive to fluctuations in $\mathcal{M}$ [as shown in Fig.~\ref{Fig_Fluctuations}(b)].

In addition to the relative sensitivity of the different memory protocols, the magnitude of memory efficiency fluctuations is of practical interest. In Fig.~\ref{Fig_Fluctuations}(d) we plot the dependence of efficiency fluctuations on memory parameter fluctuations for the four points shown in Fig.~\ref{Fig_Fluctuations}(b) spanning all three physical protocols. In each case, $\sigma_\eta^\text{fluc}$ is roughly linear in $\epsilon_M$, with proportionality constants $p =$ 0.38, 0.13, and 0.09 in the absorb-then-transfer, ATS, and EIT regions, respectively. Insofar as fluctuations in memory parameters are not amplified in the resulting fluctuations in memory efficiency ($p < 1$), it can be said that all three protocols are `stable.' although the EIT and ATS protocols are significantly more stable than the absorb-then-transfer protocol.

\section{Drift and Improper Control Field Setting in Resonant $\Lambda$-type Quantum Memory}\label{DriftSec}

In this section we consider the sensitivity of $\Lambda$-type quantum memory to long-timescale drift in control field parameters $\mathcal{G}$ at fixed memory parameters $\mathcal{M}$. The following analysis equivalently provides an indicator of the region of control field phase space where acceptable memory performance is achievable. We perform this analysis on both control fields of Gaussian temporal envelope and control fields with arbitrarily shaped optimal temporal envelope. Analysis of Gaussian control fields is significantly less computationally expensive, and allows for one-at-a-time (OAT) and global sensitivity analysis, whereas it is only computationally feasible to perform OAT sensitivity analysis on arbitrarily shaped control fields.

\subsection{One-at-a-time (OAT) Analysis}\label{OATsec}

\begin{figure}[t]
	\centering
	\includegraphics[width=\linewidth]{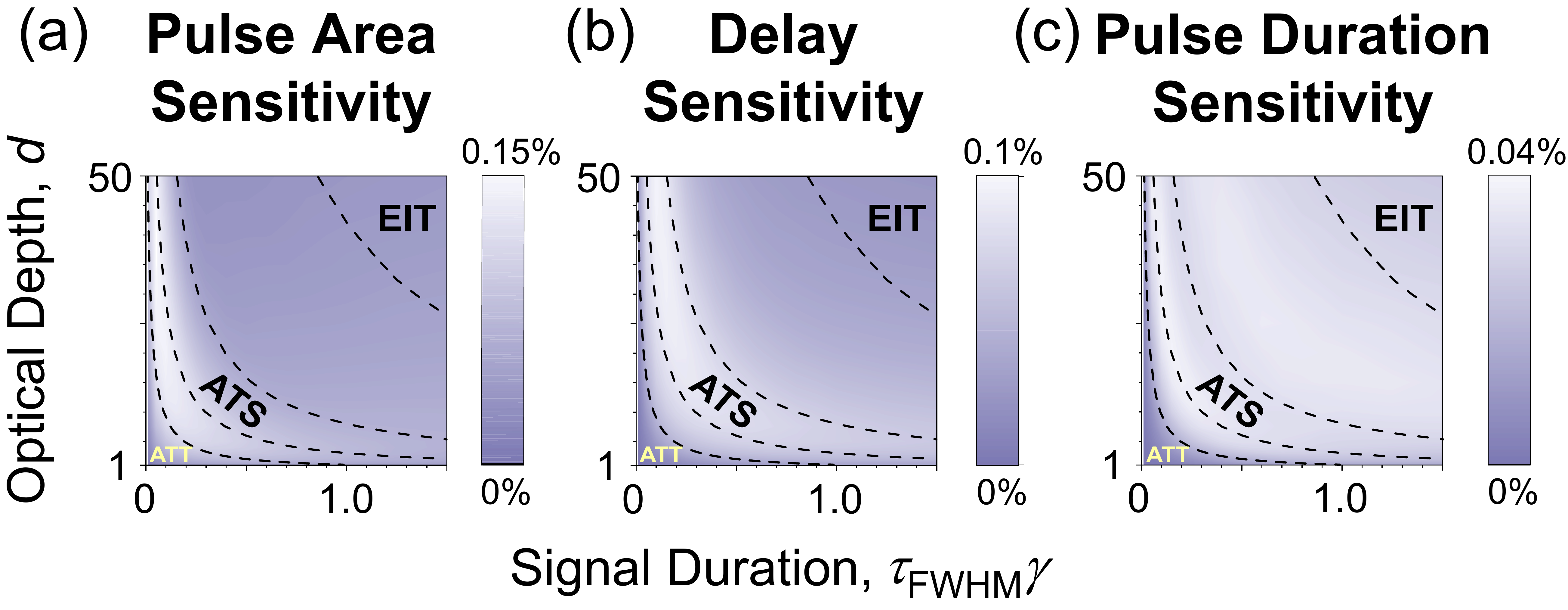}
	\caption{
	One-at-a-time sensitivity of $\Lambda$-type quantum memory to drift or improper setting of control field variables $\mathcal{G}_G$ of a Gaussian control field, as a function of memory parameters $\mathcal{M} = (d,\tau_{\text{FWHM}}\gamma)$. Regions for ATT, ATS, and EIT memory protocols are enclosed with dotted lines.}
	\label{Fig_GaussOATDrift}
\end{figure}

We first consider OAT variation of the control field parameters for a Gaussian optimized control field, $\mathcal{G}_G$. We allow for drift or improper setting of the control field pulse area, delay (relative to the signal field), and pulse duration within $\epsilon_G = 5\%$ of the optimal setpoints for any given memory parameters. 


The results are shown in Fig.~\ref{Fig_GaussOATDrift}(a)-(c). We find that sensitivity to drift or improper setting in pulse area [Fig.~\ref{Fig_GaussOATDrift}(a)] is largest in the region of $\mathcal{M}$-space below the ATS region. 
As the memory protocol in this region relies on exactly $\pi$-pulse control fields to transfer atomic population from the excited to storage state, this result agrees with physical intuition. By contrast, memory sensitivity to drift or improper setting of control field delay is largest in the ATS region [Fig.~\ref{Fig_GaussOATDrift}(b)]. The ATS protocol relies on signal and control fields that overlap in time in order to implement the requisite dynamically controlled Autler-Townes splitting \cite{ATS}, whereas the absorb-then-transfer and EIT protocols are relatively robust to improper control field delay setting. Again the ATS protocol, and the region of mixed ATS-EIT memory behavior, are most sensitive to variation in control field pulse duration [Fig.~\ref{Fig_GaussOATDrift}(c)], as changes to control pulse duration affect the dynamical Autler-Townes splitting and the effective pulse area of the control field overlapping with the signal field.

\begin{figure}[t]
	\centering
	\includegraphics[width=\linewidth]{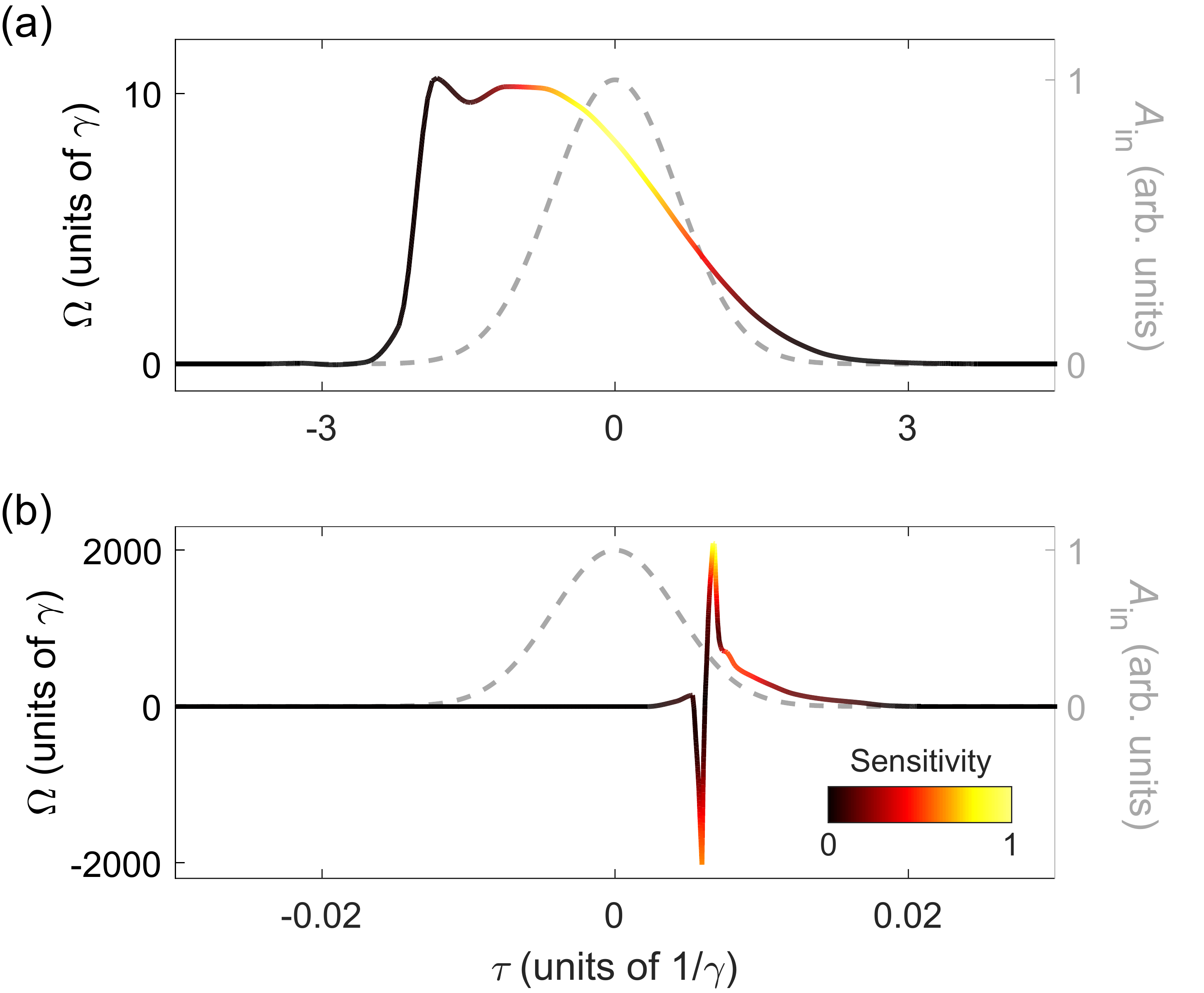}
	\caption{One-at-a-time sensitivity of $\Lambda$-type quantum memory to drift or improper setting of arbitrary optimized control field shape, $\mathcal{G}_s$, for optical depth $d=50$ in the (a) adiabatic regime, where $\tFWHM\gamma = 1.5$, and (b) non-adiabatic regime, where $\tFWHM\gamma = 0.01$. Grey dotted lines show $A_\text{in}(\tau)$. }
	\label{Fig_ShapeDrift}
\end{figure}

We also perform an OAT sensitivity analysis for arbitrarily shaped optimal control fields. For arbitrarily shaped control fields, given the large number of independent variables used to define the control field shape, computation of the shape sensitivity over a large range of optical depths and signal pulse durations is computationally intractable. Instead, we pick two pulse durations---$\tFWHM = 1.5$ and $\tFWHM = 0.01$---at an optical depth of $d=50$, corresponding to adiabatic and non-adiabatic memory conditions, respectively. We parameterize each control field shape with 51 and 135 independent spline points spaced on a Chebyshev grid for $\tFWHM = 1.5$ and $\tFWHM = 0.01$, respectively. Each point along the optimal control field shape is allowed to vary by 5\% of its optimal setpoint, as shown in Fig.~\ref{Fig_Gauss_v_Shape}(b), and the resulting variance and standard deviation in memory efficiency is recorded. The results of this OAT analysis are shown as a heat map in Fig.~\ref{Fig_ShapeDrift}. 
For $\tFWHM=1.5$ [Fig.~\ref{Fig_ShapeDrift}(a)], we find the falling edge of the typical EIT-like control field shape to be most sensitive to drift or improper setting, in agreement with Ref.~\cite{Guo19}. 
Note that larger $\tau$ on the x-axis of Fig.~\ref{Fig_ShapeDrift} corresponds to later times in the co-moving frame. For $\tFWHM=0.01$, the optimal control field shape shows characteristic non-adiabatic ringing, and it is the points on the shape with the largest amplitude that show the largest sensitivity, likely due to the fact that changes in these points cause the greatest change in the control field pulse area, which must remain close to $\pi$ in this regime. The storage efficiencies for these two optimized pulse shapes are 95.2\% and 58.7\%, for $\tFWHM = 1.5$ and $\tFWHM = 0.01$.

\subsection{Sobol' Analysis}\label{Sobolsec}

\begin{figure*}[t]
	\centering
	\includegraphics[width=0.8\linewidth]{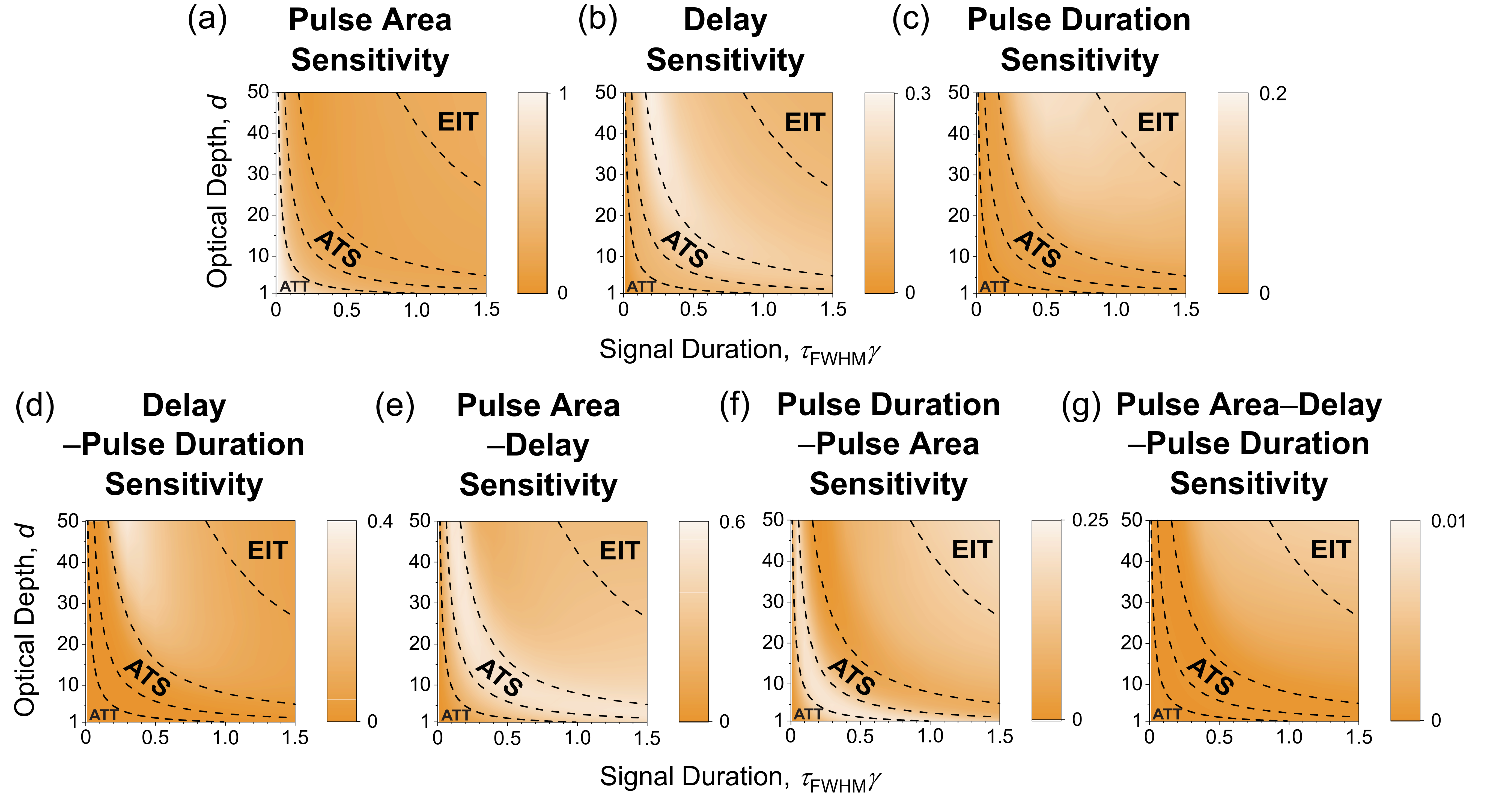}
	\caption{Sobol' sensitivity indices for $\Lambda$-type quantum memory subject to drift or improper setting of control field variables $\mathcal{G}_G$ of a Gaussian control field, as a function of memory parameters $\mathcal{M} = (d,\tau_{\text{FWHM}}\gamma)$. Regions for ATT, ATS, and EIT memory protocols are enclosed with dotted lines.}
	\label{Fig_SobolDrift}
\end{figure*}

The OAT analysis above provides rapid cross-sectional information on quantum memory performance around the optimal control field setpoints and ignores any correlations between control field parameters. Given the physical descriptions of the three resonant quantum memory protocols, we expect correlations between control field parameters in both the ATS and EIT regimes. In the ATS regime, quantum storage is accomplished with a $2\pi$ pulse-area control field which overlaps in time with the incident signal field; if, for example, the control field pulse duration is erroneously set to be too long, we expect that this error can be compensated for---with minimal effect on storage efficiency---with a larger pulse area, such that the net pulse area over the temporal extent of the signal field remains $2\pi$. In the EIT regime, quantum storage is implemented with a control field that opens a transparency window at the signal wavelength and then slowly closes this transparency window as the signal field is compressed and propagates through the medium [see, e.g., Fig.~\ref{Fig_ShapeDrift}(a)]. In the EIT case, we expect correlations between all three Gaussian control field parameters, as it is the slope of the falling edge of the control field which is most important. Errors or drift in control field delay, for example, may be compensated for with larger or smaller pulse area and pulse duration, depending on the sign of the change in delay. (If the change in delay is positive, i.e., the control field shifts closer to the signal field, a reduced pulse area and pulse duration will maintain the same slope of the control field as it closes the transparency window around the signal field.) We expect no parameter correlations in the absorb-then-transfer regime, as this protocol relies on $\pi$ pulse-area control fields that arrive after the signal field. Errors or drift in pulse area, for example, cannot be compensated for with changes to control field pulse duration or delay. Note that in the definition of control field pulse area, we account for a given pulse duration; therefore, changes to pulse duration in our model do not directly affect pulse area, and vice versa \cite{GaussianPaper}. 

In order to probe these correlations, to determine the \textit{global} sensitivity of each regime, and to determine which parameters to tune in order to compensate for drift in a given parameter, we use the first and higher-order Sobol' analysis described in Sec.~\ref{DefSec}. The results of this analysis are shown in Fig.~\ref{Fig_SobolDrift}. Importantly, Sobol' sensitivity analysis also allows for computing single-parameter sensitivities [Eqs.~\eqref{Vi}-\eqref{Si}], which are shown for Gaussian control field parameters (pulse area, delay, and pulse duration) in Fig.~\ref{Fig_SobolDrift}(a)-(c). The results of this single-parameter sensitivity calculation largely agree with the OAT analysis of Sec.~\ref{OATsec} and Fig.~\ref{Fig_GaussOATDrift}, but are in principle more reliable. The true advantage to this analysis however is in the two- and three-parameter sensitivities shown in Fig.~\ref{Fig_SobolDrift}(d)-(f) and Fig.~\ref{Fig_SobolDrift}(g), respectively. In Fig.~\ref{Fig_SobolDrift}(d), 
we calculate the second-order Sobol' sensitivity index for varying control field delay and pulse duration. We observe the largest sensitivity in the mixed ATS-EIT region, implying that these parameters are tightly correlated in this region of memory parameters. Fig.~\ref{Fig_SobolDrift}(e) shows significant correlations between pulse area and control field delay in the ATS regime, which we can interpret physically as follows: any drift or deviation in control field delay away from 0 in the ATS regime can be compensated for with a larger control field pulse area, such that the effective pulse area overlapping with the temporal extent of the signal field---which controls the dynamical Autler-Townes splitting---is still 2$\pi$. Fig.~\ref{Fig_SobolDrift}(f) shows a similar correlation between pulse duration and pulse area in the mixed ATT-ATS region, where presumably it is the effective pulse area overlapping with the signal field that is most important for high efficiency quantum storage. Fig.~\ref{Fig_SobolDrift}(f) also shows some correlation between control pulse duration and pulse area in the EIT regime. A similar behavior can be seen in the 
three-parameter sensitivities shown in Fig.~\ref{Fig_SobolDrift}(g), although the magnitude of the sensitivity index is smaller. We explain both of these correlations along the same lines discussed in Sec.~\ref{OATsec} 
for Fig.~\ref{Fig_GaussOATDrift}(b)---that the physically important part of the control field in the EIT regime is the trailing edge that closes the transparency window around the signal field, and therefore any deviations in pulse area, delay, or duration can be compensated for with the remaining degrees of freedom to ensure that the slope of the falling edge of the control field remains close to the same. The comparison between Fig.~\ref{Fig_SobolDrift}(f) and (g) show however that EIT control pulse duration and pulse area are more tightly correlated than all three parameters together. This means that if there is drift in control field pulse area in the EIT regime, for example, most of the corresponding change in memory efficiency can be compensated for by tuning the the control pulse duration (and vice versa).

\section{Conclusion}

In this work we have presented a general framework for evaluating the sensitivity of $\Lambda$-type optical quantum memory to fluctuations and drift (or improper setting) of memory and control field parameters. We have applied this framework for the case of non-negligible memory parameter fluctuations, where we have found that the region of memory parameter space corresponding to the absorb-then-transfer protocol is most sensitive, yet for all memory parameters, $\Lambda$-type optical quantum memory is stable insofar as the resulting fluctuations in memory efficiency are always smaller than the magnitude of the fluctuations in memory parameters. Further we have considered the case of drift or improper setting of control field parameters in both the case of Gaussian control fields and arbitrarily shaped control fields. The collapse of the $N$-dimensional parameter space in the case of full arbitrary shape-based optimization to just 3 physically instructive dimensions in the Gaussian case allows for a physical interpretation of memory sensitivity, as well as more a sophisticated Sobol' analysis of correlations between control field parameters.

To determine which analysis is most relevant for a given quantum memory experiment, the experimentalist must measure or estimate the fluctuations in memory parameters, $\epsilon_M$, and drift in control field parameters, $\epsilon_G$, to determine which contribution dominates. If $\epsilon_G$ is dominant, the experimentalist must then determine if a single control field parameter is drifting, in which case the one-at-a-time (OAT) analysis of Sec.~\ref{OATsec} is sufficient, or if all control field parameters are drifting by close to the same fractional amount, in which case the Sobol' analysis of Sec.~\ref{Sobolsec} is necessary. If $\epsilon_M$ and $\epsilon_G$ are of similar magnitude, both analyses may be necessary to accurately estimate the memory sensitivity and resulting fluctuations and drift of the memory efficiency. 

We note that many alternative calculations to those presented here are also possible. For example, if shot-to-shot fluctuations exist in the control field parameters instead of the memory parameters, the same analysis of Sec.~\ref{FlucSec} may be applied, substituting $\mathcal{G}$ for $\mathcal{M}$. Variations in two-photon or single-photon detuning as well as chirp in the optical fields can also be implemented in a straightforward fashion.


\ \\

\section*{Acknowledgements}

We gratefully acknowledge helpful discussion provided by Yujie Zhang, Kathleen Oolman, Dongbeom Kim, Donny Pearson, and Elizabeth Goldschmidt; support from NSF Grants No. 1640968, No. 1806572, No. 1839177, and No. 1936321; and support from NSF Award DMR1747426. This work made use of the Illinois Campus Cluster, a computing resource that is operated by the Illinois Campus
Cluster Program (ICCP) in conjunction with the National Center for Supercomputing Applications (NCSA) and which is supported by funds from the University of Illinois Urbana-Champaign.

\bibliographystyle{apsrev4-1}
\providecommand{\noopsort}[1]{}\providecommand{\singleletter}[1]{#1}%
\end{document}